\documentstyle{article}      
\begin{document}      
\title{Infinite-Dimensional Linear Dynamical \\  
            Systems with Chaoticity   
\thanks{Research supported by the National Natural Science 
        Foundation of China Grant No.19572075 for X. Fu, 
        and by the National Science Foundation of USA Grant DMS-9704345
        for J. Duan.}}   
\author{{Xin-Chu FU} \\ 
	{\small Mathematics Institute } \\
	{\small University of Warwick } \\
	{\small Coventry CV4 7AL, UK} \\
     		{\small	and  } \\
          {\small  Wuhan Institute of Physics and Mathematics  } \\
          {\small  The Chinese Academy of Sciences  }\\ 
          {\small  P. O. Box 71010, Wuhan 430071, CHINA }  \\  \\
        {Jinqiao DUAN} \\ 
          {\small Department of Mathematical Sciences }\\ 
          {\small Clemson University  }\\
          {\small Clemson, SC 29634, USA } }  
       
\date{22 December, 1997}
\maketitle

\noindent{\bf Abstract:}  The authors present two results on
infinite-dimensional linear dynamical systems with chaoticity.  One is about the
chaoticity of the backward shift map in the space of infinite sequences on a
general Fr\'{e}chet space.  The other is about the chaoticity of a translation
map in the space of real continuous functions.  The chaos is shown in the senses
of both Li-Yorke and Wiggins.  Treating dimensions as freedoms, the
two results imply that in the case of an infinite number of freedoms,  a
system may exhibit complexity even when the action is linear.  Finally, the
authors discuss physical applications of infinite-dimensional linear chaotic
dynamical systems.

\bigbreak
 
\bigbreak
\noindent{\bf Running Head:}  \ \  Linear Chaotic Systems

\newpage 

\section*{\bf 1. Introduction}

Since the mid-1970's, the experimental and theoretic research on chaotic  
phenomena and complexity of nonlinear systems has made great progress. 
A new science called nonlinear science has emerged. It is widely 
accepted  that the nonlinearity is the source which leads a system to chaos. 
And in general, it is also thought that only nonlinear systems may be chaotic.

But what is the intrinsic connection between nonlinearity and chaoticity? Is 
there a linear system with chaoticity? We have discovered that the other 
factor, the dimension of the phase 
space of the system, plays an important role.

The relation among linearity, nonlinearity, dimensions and 
chaoticity may be roughly expressed as follows:

(A) Finite dimensions and linearity implies ``nonchaoticity";

(B) Finite dimensions and  chaoticity implies ``nonlinearity";

(C) Chaoticity implies  ``finite dimensions and nonlinearity" or 
``infinite dimensions".

We thank the referees for pointing out relevant papers by
Protopopescu, MacCluer, Gulisashvili, Chan and Shapiro  (see \cite{P}, 
\cite{CS}, \cite{PA}, \cite{M}, and \cite{GM}).
We constructed an infinite-dimensional linear chaotic system in 1990, 
and earlier related results were presented  
during 1990-1993 ( see \cite{Fu2}, \cite{Fu1}, and \cite{Fu3} ).
Godefroy and Shapiro (\cite{GS}) showed that a class of linear operators are chaotic.
In particular, for the backward shift map $B$ in separable Hilbert spaces, the
map $\lambda B$ with $ |\lambda | > 1 $ is chaotic, as also discussed in 
Protopopescu
\cite{P} and MacCluer \cite{M}.  MacCluer \cite{M} further showed that continuous
semigroup $e^{tB}$, with $B$ the backward shift map 
in separable Hilbert spaces,
is chaotic.  He also showed that this is 
true for some other operators which are
infinite series in terms of B.

Godefroy and Shapiro \cite{GS} showed that the translation operator in the Fr\'{e}chet
space of complex entire functions is chaotic, and this generalized Birkhoff's
result \cite{Birkhoff}.  Chan and Shapiro (\cite{CS}) further generalized this work to the space
of slowly growing entire functions.

Linear chaotic infinite dimensional systems may be meaningful models for
physical problems.  For example, Protopopescu and Azmy \cite{PA} 
considered a infinite
dimensional system of linear differential equations in the modeling of motions
of particles of internal energy, and showed that it is chaotic.  
MacCluer \cite{M}
also pointed out that some linear models are useful for in modeling tool
chatter, and modal interaction in structures.  Gulisashvili and MacCluer 
\cite{GM}
showed that a linear quantum harmonic oscillator is chaotic by using a result of
Godefroy and Shapiro \cite{GS}

In all these cited papers, the chaos is in the sense of 
Wiggins \cite{Wiggins} or Devaney \cite{Devaney} in terms
of topological transitivity (dense orbits), sensitive dependence on 
initial conditions, and sometimes, dense periodic points.

In this paper,  
we will present two theorems on infinite-dimensional linear dynamical
systems with chaoticity. In Section 2  we consider the backward 
shift map in the space of infinite sequences of elements from
 a general Fr\'{e}chet space, and show that this backward
shift map {\em itself} is chaotic in both senses of Wiggins, and
of Li-Yorke \cite{LY} in terms of orbit separation of nonwandering 
non-periodic points.
In  Section 3 we  show that the translation map in the space of real
continuous functions is chaotic in both senses of Wiggins, and
of Li-Yorke. 

Note that we  have only proved
the backward shift itself is chaotic in the space of infinite sequences
with elements from any Fr\'{e}chet spaces, in both senses of Wiggins and of
Li-Yorke.
Godefroy and Shapiro's result (\cite{GS}) implies that the backward shift
multiplied by a constant exceed $1$ is chaotic in the sense of Wiggins.

Our work on the translation map is posed in 
 the space of real continuous functions. This space may be  more natural
 for considering  physical problems modeled by (linear) differential 
 equations, than the space of complex entire functions as used by
 Godefroy and Shapiro \cite{GS}.  Again, our work is in both senses
 of Wiggins and of Li-Yorke.
 
There are a few definitions of ``chaos".  Although the definition in Wiggins
\cite{Wiggins} seems a popular one, other definitions \cite{LY}, \cite{KS},
which capture or describe various different dynamical behavior of a system, are
proposed and used in nonlinear science.  Li-Yorke's definition of chaos does
characterize the complexity of dynamical systems.  For example, for
one-dimensional dynamical systems (the iteration of continuous maps on
intervals), Li-Yorke's chaos is equivalent to having positive topological
entropy.  The same conclusion holds for subshifts of finite type.  This
motivates us to study chaos of the backward shift map and translation map in the
sense of Li-Yorke.  Our results provide tools to analyse chaotic behavior,
especially Li-Yorke type, of infinite-dimensional linear physical systems.  In
Section 4, we will briefly discuss two linear physical systems which are chaotic
in the senses of both Wiggins and Li-Yorke.

\section*{\bf 2. Construction  of a Linear Chaotic System: Discrete Case}

Before   discussing  infinite dimensional linear chaotic systems,  
we recall two   definitions of chaos (see \cite{LY}, \cite{Devaney},
 \cite{Wiggins}, \cite{Zhou}).

\noindent{\bf Definition 1.}  Let $(M,\rho )$ be a metric space, 
$F: M\rightarrow M $ be continuous. The discrete dynamical system 
$(M,F)$ is called chaotic in the sense of Li-Yorke if there exists an 
uncountable subset $S$ of nonwandering and non-periodic points  such that 
the following conditions hold, 
$$
\left.\begin{array}{l} 
(i)\ \limsup\limits_{n\rightarrow +\infty }\rho (F^n(x), F^n(y))>0\\
(ii)\ \liminf\limits_{n\rightarrow +\infty } \rho (F^n(x),F^n(y))=0
\end{array}\right.,   \;\;\;  \forall x,y\in S, x\not= y.
$$
 
\it Remark 1.\rm\ \ Note that the Li-Yorke type chaos is characterized
via orbit separation. The original characterization of chaos in Li-Yorke's  
theorem (\cite{LY}) is via three conditions. The third one is :
$$
(iii)\ \limsup\limits_{n\rightarrow +\infty }\rho(F^n (x),F^n(p))>0, 
$$
for $\forall x \in S$ and for any periodic   point $p$.
This condition means that  there exist no asymptotically periodic points in 
$S$. From the conditions (i) and (ii) in Definition 1, it
can be shown that $S$ contains at 
most one asymptotically periodic point. So the condition (iii) is 
not essential and then removable. This observation seems
not well-known and we briefly prove it here (see \cite{Zhou} and 
\cite{Fu6} for detail). Namely, we show that,
under conditions (i) and (ii), $S$ contains at most one 
asymptotically periodic point. Note that asymptotically periodic points
are also nonwandering points. Suppose that $S$ contains two 
asymptotically periodic points $x_1, x_2$, i.e., there exist
periodic points $p_1, p_2$ (which may or may not be the same) of $F$, 
such that
$$
\lim_{n\rightarrow +\infty } \rho(F^n (x_i),F^n(p_i))=0, i=1,2.
$$
Note that 
\begin{eqnarray*}
&  & \rho(F^n(p_1), F^n(p_2))-\rho(F^n(p_1),F^n(x_1)) -\rho(F^n(x_2), F^n(p_2)) 
\\
& \leq & \rho(F^n(x_1), F^n(x_2))   \\
& \leq & \rho(F^n(x_1), F^n(p_1))+\rho(F^n(p_1),F^n(p_2))+\rho(F^n(p_2), 
F^n(x_2)),
	\forall n \geq 0.
\end{eqnarray*}
Hence, if $p_1=p_2$, then
$$
 \lim_{n\rightarrow +\infty } \rho (F^n(x_1), F^n(x_2)) = 0,
$$
which contradicts condition (i).
If $p_1 \neq p_2$, then 
$$
 \liminf\limits_{n\rightarrow +\infty }  \rho (F^n(x_1),F^n(x_2))>0,
$$
which contradicts condition (ii). So $S$ can   contain
at most one asymptotically periodic point, and thus condition (iii)
is not needed.
 
\noindent{\bf Definition 2.} Let $M,\rho , F$ be the same as in 
Definition 1, and $S \subseteq M $ be a compact invariant subset of $F$.
$(M,F)$ is said to be chaotic in the sense of Wiggins, if 

(a) $F$ has sensitive dependence on initial conditions on $S$; 
i,e., there exists 
$\delta >0$ such that for any $x$ in $S$ and any neighborhood $U$ of $x$, 
there exists $y \in U$ and $n\geq 0$ such that
$$
\rho (F^n(x),F^n(y))>\delta ;
$$
\indent (b) $F$ is topologically transitive on $S$; i.e.,
for any two open 
subsets $U$ and  $V$ in $S$, there exists $k>0$, such that
$$
F^k(U)\cap V\not= \O.
$$

\it Remark 2.\rm\ For the case of a flow, it is not difficult to give 
corresponding definitions as Definition 1 and 2.
In Definition 2, the case that $S$ is without interior points is
a trivial case. We do not treat this special case.

In \cite{Devaney}, Devaney adds the third condition to   
Definition 2: \\ 

\indent (c) The periodic points of $F$ are dense in $S$. \\

\it Remark 3.\rm\  The conditions (b) and (c)  imply condition (a) \cite{Banks}. But 
the importance and relationship to chaos of condition (c) should not be 
neglected (see \cite{Wiggins}).

Now we define a discrete infinite dimensional linear chaotic system. 

Let $(X,d)$ be a Fr\'{e}chet space over the complex field $C$ (i.e.,a complete 
linear metric space over $C$).  We denote $\Sigma (X)$ the space
 of functions from the nonnegative integers to 
 $X, c:{\cal Z}^+\rightarrow X$, where $c(i)$ is denoted by 
 $c_i$ and $c$ may be denoted by 
$c=(c_0, c_1,\cdots )$, and define the backward shift 
$\sigma :\Sigma (X)\rightarrow \Sigma (X)$
by $(\sigma (c))_i=c_{i+1}$. And let $\Sigma (X)$ be endowed with 
the product topology with the metric  
$$
\rho (x,y)=\sum \limits^{+\infty }_{i=0} \frac{1}{2^i} 
\frac{d(x_i,y_i)}{1+d(x_i,y_i)},
\ \ x=(x_0, x_1,\cdots ),y=(y_0,y_1, \cdots )\in \Sigma (X).
$$
This generalizes the usual symbolic dynamics system $(\Sigma^N, \sigma)$
(\cite{Wiggins}).

\indent \it Remark 4.\rm\  For a Fr\'{e}chet space $(X,d)$, 
 the metric $d$, or $d'$
which is equivalent to $d$, is translation-invariant. So the metric $\rho$ on 
$\Sigma(X)$ may be regarded as translation-invariant. 

Define the addition ``$\oplus$'' and the scalar multiplication 
``$\cdot$'' in $\Sigma (X)$ as follows:
$$
\left.\begin{array}{l} 
(x \oplus  y)_i=x_i+y_i\\
(\alpha \cdot x)_i=\alpha x_i
\end{array}\right. ,\ \  x,y\in \Sigma (X),\alpha \in C, i=0,1,\cdots ,
$$
then it is not difficult to verify that $(\Sigma (X),\rho )$ is a 
Fr\'{e}chet space over $C$. From
$$
\rho (\sigma (x),\sigma (y))\leq 2\rho (x,y), 
$$
$\sigma $ is continuous.$(\Sigma (X),\sigma)$ is the general symbolic 
dynamics system (see \cite{Fu6} chapter 13 or \cite{Fu5}).


It may be verified that
$$
\sigma (\alpha \cdot x\oplus \beta \cdot y)=\alpha \cdot \sigma (x) \oplus  
\beta \cdot \sigma (y),\ \ \forall x,y\in \Sigma (X),\forall \alpha , 
\beta \in C,
$$
that is, $\sigma: \Sigma (X)\rightarrow \Sigma (X)$ is a linear map, and  
$(\Sigma (X),\sigma)$ a linear dynamical system.

For the orbit $\{\sigma ^n(x),n\geq 0\}$ from $x$ and the orbit 
$\{\sigma ^n(y),n\geq 0\}$ from $y$, consider their linear combinations.  
We have
$$
\alpha \cdot \sigma ^n(x)\oplus \beta \cdot \sigma ^n(y)=\sigma^n(\alpha 
\cdot x\oplus \beta \cdot y),
$$
i.e., $\{\alpha \cdot \sigma ^n(x)\oplus \beta \cdot \sigma ^n(y), n\geq 0\}$
is the orbit from $\alpha \cdot x\oplus\beta \cdot y$. So, the system 
$(\Sigma (X),\sigma )$ satisfies the orbit superposition principle .

When $X$ is nontrivial, $\Sigma (X)$ is infinite-dimensional.


We denote $x=(x_0, x_1,\cdots )^T \in \Sigma (X)$ as an infinite-dimensional 
column vector. Define the infinite matrix
$$
A=(a_{ij})_{i,j=0,1,\cdots ,+\infty },
$$
where
$$
a_{ij}=\left\{\begin{array}{l} 1,\ \ j=i+1,\\
0,\ \ \hbox{otherwise.}\end{array}\right. 
$$
Then the map $\sigma$ has a matrix representation
$$
\sigma (x)=Ax,
$$
which can also be written as 
$$
\sigma (x)=(\sum \limits^{+\infty }_{k=0}a_{0k}x_k,
\sum \limits^{+\infty }_{k=0}a_{1k}x_k,\cdots ).
$$
\indent For any $\lambda \in C,\lambda$ is an eigenvalue of $\sigma$,
and there are infinitely many eigenvectors to an eigenvalue $\lambda$,
$$
x=(x_0, \lambda x_0,\lambda ^2x_0,\cdots ,\lambda ^kx_0, \cdots ),
\ \  \forall x_0\not= 0.
$$
\indent The following proposition is obvious:

\noindent{\bf Proposition 2.1.} \it\  (1) $\forall \lambda \in C, \lambda$ is an 
eigenvalue of $\sigma$, and the corresponding eigenspace is isomorphic to $X$.
(2) The spectrum of $\sigma$ consists of eigenvalues.  \rm

Many familiar operators have finite or infinite countable
number of eigenvalues in
discrete distribution.  But the map $\sigma$ here has {\em uncountable} 
eigenvalues in
continuum distribution.  This is an indication that 
the system $(\Sigma (X),\sigma )$  may have complicated properties.  
The following two theorems justify this observation. The first result,
Theorem 2.2, was included here  to point out that the subshifts of
$(\Sigma (X),\sigma )$ can be used as {\em models} of other
continuous  maps in infinite dimensional metric spaces.

\noindent{\bf Theorem 2.2}(X.-C. Fu \cite{Fu2}). \it \  Let $(X,d)$ be a metric space. 
For any 
continuous  map $\tau:X\rightarrow X$, there exists a 
subshift $(\Sigma _\tau , \sigma _\tau )$ 
of $(\Sigma (X),\sigma )$, such that $\tau$ is topologically 
conjugate to $\sigma _\tau$,where $\sigma_\tau$ is $\sigma$ restricted to
$\Sigma_\tau$.\rm

The above theorem shows the importance of subshifts of $(\Sigma (X), \sigma)$. 
And meanwhile it reveals the extraordinarily plentiful structure of subshifts 
in $(\Sigma (X)$, $\sigma)$. Such a system $(\Sigma (X),\sigma)$, probably, 
should have chaotic properties. Indeed, we have:

\noindent{\bf Theorem 2.3.}\it \ \ Suppose $(X,d)$ is a nontrivial 
separable Fr\'{e}chet space; 
then the infinite-dimensional linear dynamical system 
$(\Sigma (X), \sigma)$ is  chaotic in the senses of both Li-Yorke and Wiggins. 
\rm

\it Proof.\rm \ \ 
We remark that our result on the chaoticity of the backward shift   
is constructive and direct, and it appears that this result does not
follow directly from Corollary 1.5 in Godefroy and Shapiro \cite{GS} as used
in chaos proofs of \cite{P}, \cite{CS}, \cite{PA}, \cite{M}, and \cite{GM}. 

We first prove that $\sigma$ is chaotic in the sense of   Li-Yorke,
that is, we will construct an uncountable set satisfying the conditions 
in Definition 1.

Choose arbitrarily $a,b\in X,a\not= b$. And take an onto 
map
$$
\alpha :(0,1)\rightarrow X-\{a\}.
$$
$\forall r\in (0,1)$, define $x^r=(x^r_0,x^r_1, \cdots )\in \Sigma (X)$ as 
follows:
$$
\left\{\begin{array}{l}
x^r_0=b\\
x^{^r}_{{_{k^2}}}=\left\{\begin{array}{l} a,\ \ \varphi (k,r)=1\\
b,\ \ \varphi (k,r)=0\end{array}\right. \\
x^r_l=\left\{\begin{array}{l} b,\ \ k^2+1\leq l\leq (k+1)^2-2\\
\alpha (r),\ \ l=(k+1)^2-1\end{array}\right. \end{array}\right. 
$$
where $k=1,2,\cdots ,\varphi (k,r)=[kr]-[(k-1)r]$ (may be called the Zhou 
function (see \cite{Fu6} and \cite{Zhou})),$[\cdot]$ denotes the integral part of a real number. 
Then $x^r=(x^r_0,x^r_1,\cdots )\in \Sigma (X)$.

We denote $\Omega(\sigma)$ as the set of nonwandering points of
$\sigma$ and $P(\sigma)$ the set of periodic points.
Let $S=\{x^r|r\in (0,1)\}$; then $S\subseteq\Omega(\sigma)-P(\sigma)$
(Note that $\Omega(\sigma)=\Sigma(X)$).

Denote by $P(x^r,k)$ the number of $x^r_l$'s which satisfy 
$x^r_l=a,0\leq l\leq k$. Then
$$
[\sqrt{k}r]\leq P(x^r,k)\leq [(\sqrt{k}+1)r],
$$
this implies
$$
\lim\limits_{k\rightarrow +\infty }\frac{P(x^r,k)}{\sqrt{k}}=r.  \eqno(*)
$$
When $r_1, r_2\in (0,1)$ and $r_1\not= r_2$, we have $x^{r_1}\not= x^{r_2}$
from $(*)$. So $S$ is an uncountable subset in $\Sigma (X)$.

From the construction of $x^r$, only the entries of type $x^{r}_{k^2}$  
may take the value $a$. Therefore, $\forall r_1,r_2\in (0,1), r_1\not=r_2$, 
there exist infinite number of positive integers $k_n,n=1,2,\cdots$,
such that $x^{r_1}_{k^2_n}\not= x^{r_2}_{k^2_n}$, due to $(*)$. 
And the entries of the type 
$x^{r}_{k^2}$ only take the value $a$ or $b$, so
$$
\begin{array}{c}
\limsup\limits_{n\rightarrow +\infty }\rho (\sigma ^n(x^{r_1}), 
\sigma ^n(x^{r_2}))\geq \limsup\limits_{n\rightarrow +\infty}\rho 
(\sigma ^{k^2_n}(x^{r_1}), \sigma ^{k^2_n}(x^{r_2}))\\
\geq \frac{d(a,b)}{1+d(a,b)}>0.
\end{array}
$$
\indent When $k^2+1\leq l\leq (k+1)^2-2,x^{r_1}_l=x^{r_2}_l=b,\forall r_1, 
r_2\in (0,1)$. And $(k+1)^2-2-(k^2+1)=2(k-1)$, so we have
$$
\begin{array}{c}
\liminf\limits_{n\rightarrow +\infty }\rho(\sigma ^n(x^{r_1}), 
\sigma ^n(x^{r_2}))\leq \liminf\limits_{k\rightarrow +\infty } \rho (\sigma 
^{k^2+1}(x^{r_1}), \sigma ^{k^2+1}(x^{r_2}))\\
\leq \liminf\limits_{k\rightarrow +\infty } \frac{1}{2^{2(k-1)}}=0.
\end{array}
$$
\indent Therefore we have proven that
$$
\limsup\limits_{n\rightarrow +\infty }\rho (\sigma ^n(x), \sigma ^n(y))
>0,\ \ \forall x,y\in S\  with\  x\not= y,
$$
$$
\liminf\limits_{n\rightarrow +\infty } \rho (\sigma ^n(x), \sigma ^n(y))=0, 
\ \ \forall x,y\in S.
$$
That is, $(\Sigma (X),\sigma )$ is a chaotic system in the sense of Li-Yorke.

We now prove that $\sigma$ is chaotic in the sense of Wiggins,
that is, we show that $\sigma$ has sensitive dependence and is topologically
transitive on a compact invariant set.

Choose a compact subset $K$ of $X$ with more than one points, 
and define a subset  
$\Sigma (K)$ of $\Sigma (X)$ by:

$$
	\Sigma (K) = \{x \in \Sigma (X) | x_i \in K, i=0, 1, \cdots\},
$$
then $\Sigma (K)$ is a compact subset of $\Sigma (X)$ 
invariant under $\sigma$.

To continue the proof of Theorem 2.3, we need the following claim.
 
{\em Claim:} There exists a positive constant $\delta _0$ such that
$$
  \forall a\in K,\exists b\in K, \ we \ have\ \ \ d(a,b)>\delta _0. \eqno(**)
$$

\indent Let us prove this claim.
We first consider the case of 
$\inf\limits_{a\in K}\sup\limits_{b\in K}d(a,b)=+\infty$. 
Suppose in this case there is no $\delta _0$ satisfying $(**)$.
Thus, $\forall \eta >0,\exists 
a_0\in K, \forall b\in K, d(a_0,b)\leq \eta$. This leads to 
$$
\inf\limits_{a\in K}\sup\limits_{b\in K}d(a,b)
\leq \sup\limits_{b\in K}d(a_0, b)
\leq \eta <+\infty .
$$ 
which contradicts with $\inf\limits_{a\in K}\sup\limits_{b\in K}d(a,b)=+\infty$.
Hence the claim is proved in this case. 

Now we consider the case of
$\inf\limits_{a\in K}\sup\limits_{b\in K}d(a,b)\not= +\infty$.
we take $\delta _0=\frac{1}{2}\inf\limits_{a\in K}\sup\limits_{b\in K}d(a,b)$.
Then there exist $a_0,b_0\in K$, such that $d(a_0,b_0)>0$.
Thus $\forall a\in K$, we have
$$
\sup\limits_{b\in K} d(a,b)\geq\max\{d(a,a_0),d(a,b_0)\}\geq 
\frac{d(a,a_0)+d(a,b_0)}{2}\geq \frac{1}{2}d(a_0, b_0)>0, 
$$
and
$$
\inf\limits_{a\in K}\sup\limits_{b\in K}d(a,b)\geq \frac{1}{2}d(a_0, b_0)>0.
$$
So $0<\delta _0<+\infty$. Therefore,
$$
\delta _0<\inf\limits_{a\in K}\sup\limits_{b\in K}d(a,b)\leq\sup\limits_
{b\in K}d(a,b),\forall a\in K.
$$
So $\forall a\in K,\exists b\in K$, such that $d(a,b)>\delta _0$.
Therefore, the constant $\delta_0$ satisfying $(**)$ really exists.
This proves the claim $(**)$.

We now continue the proof of Theorem 2.3. Let $\delta=\delta_0/(1+\delta _0)$; then for any $x\in \Sigma(K)$ and
its any neighborhood $U$  in $\Sigma(K)$, take
$$
y^{(k)}=(x_0, x_1\cdots, x_k, y_{k+1}, \cdots),
$$
where $y_{k+1}\in K$ satisfies $d(x_{k+1},y_{k+1})>\delta _0$. When $k$ is big 
enough, we have $y^{(k)}\in U$ and 
$$
\rho(\sigma^{k+1}(x),\sigma ^{k+1}(y^{(k)}))=\frac{d(x_{k+1},y_{k+1})}{1+
d(x_{k+1},y_{k+1})}+\cdots >\frac{\delta _0}{1+\delta _0}=\delta .
$$
i.e., $\sigma$ has sensitive dependence on initial conditions 
on $\Sigma(K)$.

Because $X$ is separable and thus $K$ is separable as well, 
there exists a countable subset $A\subseteq K$,
such that the closure of $A$ is $K$.

Denote $A=\{a_0, a_1, \cdots ,a_k,\cdots \}$ and $A_N=\{a_0, a_1, \cdots,a_N\},
N\geq 1$. Let $M^N_{pq}$ be the $q$-th arrangement according to the 
lexicographic order of all repeatable words of length $p$ in the elements in 
$A_N, N\geq 1,1\leq q\leq N^p,p=1,2,\cdots$. There are countable infinite 
elements $M^N_{pq}$. So we can arrange the elements $M^N_{pq}$   
and denote them by new symbols: $B_1, B_2, \cdots,B_k,
\cdots$. We form a symbol sequence $b$ by putting together all the 
words $B_i, i=1,2,\cdots$.

\indent $\forall \varepsilon >0,\forall x=(x_0, x_1,\cdots ,x_k,\cdots )\in 
\Sigma (K)$, for $x_k\in K$, there exist $y_k\in A$, such that
$$
d(x_k,y_k)<\frac{\varepsilon }{4},\ \ k=0,1,\cdots .
$$
Take $y=(y_0, y_1,\cdots ,y_k,\cdots)$; then $\forall m>0$, by the construction 
of $b$, there is $k\geq 0$, such that the first $m+1$ entries of $y$
 and $\sigma ^k(b)$ agree, so
$$
\rho (y,\sigma ^k(b))\leq \frac{1}{2^m}
$$
i.e., the orbit $\{\sigma ^n(b),n=0,1, \cdots\}$ can be arbitrarily close to 
$y$. So there exists $n\geq 0$, such that
$$
\rho (y,\sigma ^n(b))<\frac{\varepsilon }{2}.
$$
Therefore
$$
\rho (x,\sigma ^n(b))\leq \rho (x,y)+\rho (y,\sigma ^n(b))<\sum \limits
^{+\infty }_{i=0}\frac{1}{2^i}\frac{\varepsilon }{4+\varepsilon }+\frac
{\varepsilon }{2}<\varepsilon.
$$
So,  $\{\sigma^n(b),n=0,1,\cdots\}$ is a dense orbit in $\Sigma(K)$. 

Hence, for any nonempty open subsets $U$ and $V$ 
in $\Sigma(K)$, there exist  $n_2>n_1>0$,
such that $\sigma^{n_1}(b) \in U$ and $\sigma^{n_2}(b)\in V$.
Hence $\sigma^{n_2 - n_1} (U) \cap  V $  is not empty,
and  $\sigma$ is topologically transitive on $\Sigma(K)$.

Thus, $(\Sigma(X), \sigma)$ is a chaotic system in the sense of  Wiggins.

The proof of Theorem 2.3 is complete.

\section*{\bf 3. Construction of a Linear Chaotic System: Continuous Case}

Let $B = C^0(R)$, the space of real continuous functions on real axis $R$.  
With the metric 
$$
\rho (f,g)=\sup\limits_{a>0}\min\{\max\limits_{|x|\leq a}|f(x)-g(x)|,
\frac{1}{a}\},
$$
$B$ is a complete separable metric space.
Moreover, $\rho$ is translation-invariant.

The addition and the scalar multiplication in $B$ are defined in usual 
fashion, i.e.,
$$
\left.\begin{array}{l} (f\oplus g)(x)=f(x)+g(x)\\
(\alpha \cdot f)(x)=\alpha f(x)\end{array}\right. ,\ \ \ \ \ \ \forall f,g\in 
B,\alpha ,x \in R
$$
then $B$ is an infinite-dimensional separable Fr\'{e}chet space.

Define a map $b: B\times R\rightarrow B$ as follows:
$$
\left.\begin{array}{l} b(f,t)=f_t\\
f_t(x)=f(x+t)\end{array}\right. ,\ \ \ \ \ \ \ \ \ \forall t,x\in R,
$$
then it may be verified that $b: B\times R \rightarrow B$
 is a flow in the infinite-dimensional phase space $B$. This dynamical system, 
 denoted by $B_t$, is called the Biebutov system \cite{Zhang}.

It may be verified that 
$$
b(\alpha\cdot f\oplus\beta\cdot g,t)=\alpha \cdot b(f,t)\oplus \beta \cdot 
b(g,t),
\ \ \forall f,g\in B,\forall \alpha , \beta , t\in R,
$$
so $B_t$ is a linear system. And the above equality indicates that the 
linear combination $\alpha \cdot b(f,t)\oplus \beta \cdot b(g,t)$
 of the orbits from $f$ and $g$ in $B$ is exactly the orbit from 
$\alpha \cdot f \oplus \beta \cdot g$. This means the system $B_t$ satisfies 
the solution superposition principle.

The system $B_t$ may be treated as a continuous shift system. Many topological 
dynamical systems may be embedded into $B_t$. 

Consider the discrete case, i.e., 
the restriction of $f \in C^0(R)$ on $\{0,\pm \tau ,\pm 2\tau 
, \cdots \}$, where $\tau>0$. Then $f$  corresponds to a bi-infinite 
sequence $f^\tau $:
$$
f^\tau =(\cdots ,f(-\tau ),f(0);f(\tau ), \cdots ).
$$
Let the time $t$ run discrete values $n\tau,n=0,\pm 1,\pm 2,\cdots$, then 
$$
b(f^\tau ,n\tau )=f^\tau _n,
$$
where
$$
f^\tau _n=(\cdots ,f((n-1)\tau ),f(n\tau );f((n+1)\tau ),\cdots ).
$$
Denote $\sigma =b(\cdot ,\tau )$, and 
$$
\Sigma (C^0(R))=\{ f^\tau _n|n\geq 0,f\in C^0(R)\},
$$
then $(\Sigma (C^0(R)),\sigma)$ is  a two-sided symbolic dynamics system.

A subsystem $M_t$ of $B_t$ is constructed in \cite{Zhang}, where $M_t$ is complete and 
all orbits in $M_t$ are asymptotic.

So the system $B_t$ possesses very complex structure. The further analysis 
shows that $B_t$ has chaotic properties, i.e.,we have:

\noindent{\bf Theorem 3.1.}\it\ \ The infinite-dimensional linear dynamical 
system
$B_t$ is chaotic in the senses of both Li-Yorke and Wiggins.

\it  Proof.\rm \ \  We first prove that $B_t$ is chaotic in the 
sense of  Li-Yorke.

Construct a subset $S$ in $B$ as follows:
$$
S=\{f_{[r]}|r\in (0,1)\},
$$
$$
f_{[r]}(x)=\left\{\begin{array}{l}0,\ \ x\in 
[-(a+\frac{1}{2}),-(a-\frac{1}{2})]\cup 
[a-\frac{1}{2},a+\frac{1}{2}]\\
1,\ \ x=\pm (a\pm 1)\\
\hbox{continuous interpolation,}\ \ \ x\in (-(a+1),-(a+\frac{1}{2}))\\
\ \ \ \ \ \ \ \ \ \ \ \ \ \ \ \ \ \ \ \ \ \ \ \ \ \ \ \ \ \ \ \ \ \ \ \
\cup (-(a-\frac{1}{2}),-(a-1))\\
\ \ \ \ \ \ \ \ \ \ \ \ \ \ \ \ \ \ \ \ \ \ \ \ \ \ \ \ \ \ \ \ \ \ \ \ 
\cup (a-1,a-\frac{1}{2})\cup (a+\frac{1}{2},a+1)\\
1,\ \ \hbox{all other cases}\end{array}\right. 
$$
where $a=k^2,k's$ are positive integers satisfying $\varphi (k,r)=1$, and 
$\varphi (k,r)=[kr]-[(k-1)r]$ is the  Zhou function \cite{Zhou}.

Denote by $P(f_{[r]}, A)$ the number of intervals contained in $[-(A+
\frac{1}{2}),A+\frac{1}{2}]$ in which $f_{[r]}(x)\equiv 0$, where 
$A$ is a positive integer. Then
$$
[\sqrt{A}r]\leq P(f_{[r]},A)\leq [(\sqrt{A}+1)r],
$$
$$
\lim\limits_{A\rightarrow +\infty }\frac{P(f_{[r]},A)}{2\sqrt{A}}=r.
$$
So $f_{[r_1]}\not= f_{[r_2]}$ for $r_1\not= r_2$, and $S$ is uncountable. 
Furthermore, when $r_1\not= r_2$, there exist  an infinite number of $t_n,
n=1,2,\cdots ,t_n\rightarrow \infty (n\rightarrow +\infty )$, such that
$f_{[r_1]}( t_n ) \not=  f_{[r_2]}( t_n )$ (Otherwise, we have
$P(f_{[r_1]},A)=P (f_{[r_2]}, A)+C$, this implies $r_1=r_2$, a contradiction).
So
$$
\limsup\limits_{t\rightarrow \infty }\rho (b(f,t),b(g,t))\geq 1>0,\ \ \forall 
f,g\in S,f\not= g.
$$
\indent When $x\in [k^2+1, (k+1)^2-1]\cup [1-(k+1)^2,-k^2-1], f_{[r]}(x)=1,
\forall r\in (0,1)$. And the lengths of the intervals $[k^2+1,(k+1)^2-1]$
and  $[1-(k+1)^2,-k^2-1]$ are 
$ 2k-1\rightarrow +\infty (k\rightarrow +\infty )$.
Thus
$$
\liminf\limits_{t\rightarrow \infty } \rho (b(f,t),b(g,t))=0,
	\ \ \forall f,g\in S.
$$
\indent $\forall f\in B,a>0$, take $f^{[a]}\in B$ as follows:
$$
f^{[a]}(x)=\left\{\begin{array}{l} f(x),\ \ x\in [-a,a]\\
f(x-3a),\ \ x\in [2a,4a]\\
\hbox{continuous extension},\ \ x\in (-\infty ,-a)\cup (a,2a)\cup (4a,+\infty 
)\end{array}\right. 
$$
then 
$$
\rho (f,f^{[a]})\leq \frac{1}{a},
$$
$$
\rho (f,b(f^{[a]},3a))\leq \frac{1}{a}. 
$$

In the following, $\Omega (B_t)$ is the set of nonwandering points of $B_t$,
while $P(B_t)$ is the set  of  periodic points.
From $\Omega (B_t)=\{ f\in B|\forall U\in o(f),\exists g\in U,t>0$, such that 
$b(g,t)\in U\}$, we have $\Omega (B_t)=B$. Where $o(f)$ denotes the collection 
of neighbourhoods of $f$.
Obviously, $S\cap P(B_t)=\O$. So $S\subset \Omega (B_t)-P(B_t)$. 

Therefore, $B_t$ is chaotic in the sense of Li-Yorke.

We now prove that $B_t$ is chaotic in the 
sense of Wiggins.

$\forall f\in B,\forall a>1$, let $f^{(a)} \in B$ as follows:
$$
f^{(a)}(x)=\left\{\begin{array}{l} f(x),x\in [-a,a]\\
f(x)+1,x\in [a+1,a+3]\\
\hbox{continuous extension},\  x\in (-\infty ,-a)\cup (a,a+1)\cup (a+3,
+\infty )\end{array}\right. 
$$
then $f^{(a)}$ may enter an arbitrarily small neighbourhood of $f$ for 
$a$ big enough. However,
$$
\rho (b(f,a+2),b(f^{(a)},a+2))=1.
$$
This implies that the system $B_t$ has sensitive 
dependence on initial conditions.

From the separability of $B$, there exists a countable subset $F=\{f_{(k)
}|k=0,1,\cdots \}\subset B$, such that $F$ is dense in $B$.

For a positive integer $n$, denote by $\varphi^{n}_{k}(x)$ the continuous 
function on the interval $[-(n+1),n+1]$:
$$
\varphi  ^{n} _{k}(x)
=\left\{\begin{array}{l} f_{(k)}(x),\ \ \ \ \ \ \ \ \ x\in [-n,n]\\
0,\ \ \ \ \ \ \ \ \ x=\pm (n+1)\\
\hbox{continuous interpolation},\ \ x\in (-(n+1),-n)\cup (n,n+1)
\end{array}\right. 
$$
then $\{\varphi^{n}_{k}|n=1,2,\cdots;k=0,1, \cdots \}$ is countable; and 
$$
\{\{ \varphi^{n}_{k}(x)|x\in [-(n+1),n+1]\}|n=1,2\cdots;k=0,1,
\cdots\}
$$
is a countable collection of curve segments. Make suitable translation 
transformation to the intervals of definition of the curve segments, such 
that the new intervals of definition cover the real number axis 
$R$ without overlaps. 
Correspondingly, the new curve segments join together and form a continuous 
curve on $R$, denoted by $\{\varphi (x)|x\in R\}$. Then the orbit 
$\{b(\varphi, t)|t\in R\}$ from $\varphi$ is dense in $B$. 
 
For $D>0$, we take $K=\{f\in B | \; \; |f(x)|\leq D, \forall x\in R\}$,
then $K$ is a compact invariant subset of the flow $b: B\times R\rightarrow B$.
From the above discussion, the restricted flow
$b_1: K \times R\rightarrow K$ has sensitive dependence on initial
conditions and is topologically transitive.

So $B_t$ is a chaotic system in the sense of Wiggins. 

This completes the proof of Theorem 3.1.

\section*{\bf 4. Discussions and Applications }

As we know, finite-dimensional linear systems  are nonchaotic. Dimensions may 
be treated as freedoms. The existence of infinite-dimensional linear 
chaotic systems indicates that if there are infinite number of freedoms,then 
a system may possibly create complexity even if the map (or flow) is linear.
The discussions in this paper suggest that the deeper understanding on the 
structure of infinite-dimensional linear topological spaces and the 
properties of continuous linear maps on the spaces should also be  
an important research subject  in nonlinear science. In \cite{Fu4} some new 
concepts and results are presented, and some intrinsic connections among  
dimensions of phase space, linearity, nonlinearity and complexity of a 
system are explored.

On the other hand, not every infinite-dimensional linear dynamical system is 
chaotic. For example, the global attractor of the infinite-dimensional linear 
dynamical system $(l^2,\sigma )$ discussed in \cite{Fu4} is a one-point set 
$\{\theta\}$. So the chaoticity of an infinite-dimensional linear dynamical 
system relates to the structure of phase space of the system.
Moreover, a large class of quantum mechanical systems are not chaotic
(see, \cite{Ingraham}).
 
Gulisashvili and MacCluer (\cite{GM})
showed that the annihilation operator (a weighted shift operator)
for a linear quantum harmonic oscillator is chaotic in 
the sense of Wiggins (or Devaney). Duan, Fu, Liu and Manning (\cite{DF})
have recently used result of the current paper to prove that
this annihilation operator is also chaotic in 
the sense of Li-Yorke. An unforced 
quantum harmonic oscillator, i.e., a very small frictionless mass-spring
system, is modeled by the Schr\"{o}dinger equation
(\cite{Liboff})
$$
	i \hbar \psi_t = -\frac{\hbar^2}{2m} \psi_{xx} + \frac{k}{2} x^2\psi, 
$$ 
with wave function $\psi(x,t)$, displacement $x$, mass $m$, stiffness $k$
and Planck number $\hbar$.
The nondimensionalized stationary states in the separable Hilbert space
$X=L^2(-\infty, \infty)$ form an orthonormal basis
$$
	\psi_n(x) = e^{-x^2/2}H_n(x)/\sqrt{\sqrt{\pi} 2^n n!}, 
		n=0, 1,  \cdots,
$$ 
where 
$$
	H_n(x) = (-1)^n e^{x^2}\frac{d^n}{dx^n} e^{-x^2},
$$
is the $n$th Hermite polynomial.
The natural space for the quantum harmonic oscillator is
the Schwartz class $F$ of rapidly decreasing functions in
$X=L^2(-\infty, \infty)$ as defined later.
Gulisashvili and MacCluer (\cite{GM})
defined a linear, closed, unbounded,  weighted 
backward shift $B$ on F by
$$
B: F  \rightarrow   F,     
$$
$$
B\psi_n  \equiv   \frac1{\sqrt{2}}(x+\frac{d}{dx})\psi_n  
= \sqrt{n} \psi_{n-1}.
$$
$B$ has no resolvent set since every complex number
$\lambda$ is in the point spectrum of $B$. By using a result of Godefroy and 
Shapiro (\cite{GS}),
Gulisashvili and MacCluer (\cite{GM})
have shown that the shift operator $B$ is chaotic   
(\cite{Devaney}), namely, it has
topological transitivity (dense orbits), sensitivity to  
initial conditions (orbit divergence), and density of periodic points.

In terms of the orthonormal basis $\{\psi_n\}$, 
the Schwartz class $F$ can be written as 
(\cite{GM})
$$
F = \{ \phi \in L^2(-\infty, \infty): \phi = \sum_{n=0}^{\infty}c_n \psi_n, 
    \sum_{n=0}^{\infty}|c_n|^2 (n+1)^r<\infty, \forall r \geq 0\}.
$$
$F$ is an infinite-dimensional Fr\'{e}chet space with topology 
defined by the system of semi-norms $p_r(\cdot)$ of the form
(\cite{Yosida}) 
$$
p_r (\phi) = p_r (\sum_{n=0}^{\infty} c_n \psi_n) = 
(\sum_{n=0}^{\infty}|c_n|^2 (n+1)^r)^{1/2}, \ \ \ r\geq 0.
$$
This topology on $F$ is also given by the  metric $\rho$  
$$
\rho (\phi, \psi) = \sum_{m=0}^{\infty}2^{-m} p_m (\phi - \psi)\cdot 
             (1+p_m (\phi -\psi))^{-1}.
$$

Fix ${\theta} \in (0, 1)$ and define $\phi^{\theta} = 
\sum_{n=0}^{\infty}c_n^{\theta}\psi_n$  by
$$
\left\{\begin{array}{l} 
c^{\theta}_0=0 \\
c^{\theta}_n = \left\{\begin{array}{cl} 1/ \sqrt{n!}   & \mbox{ if $ 
n = k^2, [k{\theta}]-[(k-1){\theta}] = 1 $ }\\
0   & \mbox{otherwise ,} 
\end{array}\right. 
\end{array}\right.
$$ 
where $k=1, 2, \cdots$, and  $[\cdot]$  denotes the integer part of a real 
number.

Let  $S=\{\phi^{\theta}: {\theta}\in (0,1)\}$. By using the idea and result 
in Section 2, it is shown (\cite{DF}) that all points in $S$ are nonwandering 
and nonperiodic, and $S$ is a chaotic set for $B$, i.e., the weighted shift
operator $B$ is chaotic in the sense of Li-Yorke.

Protopopescu (\cite{P}) considered an infinite-dimensional 
linear rate equation arising in the modeling of
particle distribution in statistical mechanics, and showed that it is
chaotic in the sense of Wiggins.  Liu, Fu and Duan \cite{LF}
have recently used idea and result in Section 3 to show that this linear 
rate equation is also chaotic in 
the sense of Li-Yorke.

Duan, Fu and Lawson \cite{DF2} also discussed the chaotic trajectory 
in the quantum 
state space of an arbitrary solution to the Schr\"{o}dinger equation 
for Hydrogen atom under the repeated application of ladder operators, 
which shift the eigenfunctions $u_n$ of the Schr\"{o}dinger equation into 
$u_{n-1}$ or $u_{n+1}$. 

It would be interesting to see other linear physical models which 
are chaotic in the sense of Li-Yorke or Wiggins.

\section*{\bf Acknowledgments}

The authors would like to express their thanks to Prof. Zhang Zhi-Fen at  
Beijing University for her useful materials on the Biebutov system. They 
were grateful to the referees' helpful comments and useful suggestions
as well as pointing out papers or books by 
Protopopescu, Azmy, MacCluer, Gulisashvili, Chan and Shapiro, and Ingraham. 

\bigbreak

\end{document}